\documentclass[fleqn,10pt]{wlscirep}
\title{The interplay between excitons and trions in a monolayer of MoSe$_2$}

\author[1,*]{N.Lundt}
\author[2]{E.Cherotchenko}
\author[1]{O.Iff}
\author[4]{X.Fan}
\author[4]{Y. Shen}
\author[6]{P. Bigenwald}
\author[2,4]{A. Kavokin}
\author[1,5]{S. H\"ofling}
\author[1]{C. Schneider}

\affil[1]{Technische Physik, University of W\"urzburg, Am Hubland, D-97074 Wurzburg, Germany}
\affil[2]{Physics and Astronomy School, University of Southampton, Highfield, Southampton, SO171BJ, UK}

\affil[3]{School for Engineering of Matter, Transport, and Energy, Arizona State University, Tempe, Arizona 85287, United States}

\affil[4]{SPIN-CNR, Viale del Politecnico 1, I-00133 Rome, Italy}
\affil[5]{SUPA, School of Physics and Astronomy, University of St. Andrews, St. Andrews KY 16 9SS, United Kingdom}
\affil[6]{Institut Pascal, PHOTON-N2, Clermont Université, Blaise Pascal University,
Centre National de la Recherche Scientifique, 24 Avenue des Landais, 63177 Aubière Cedex, France}

\begin{abstract}
We study the impact of a free carrier reservoir on the optical properties of excitonic and trionic complexes in a MoSe$_2$ monolayer at cryogenic temperatures. By applying photodoping via a non-resonant pump laser the electron density can be controlled in our sample and in turn the exciton and trion densities can be tuned. We find a significant increase of the trion binding energy in the presence of an induced electron gas both in power- and in time-resolved photoluminescence spectra. This behaviour is reproduced within the original variational approach that takes into account both screening and phase space filling effects.   
\end{abstract}
\begin{document}

\flushbottom
\maketitle
%
%
\thispagestyle{empty}
\section*{Introduction}
Monolayers of transition metal dichalcogenides (TMDCs) are a close-to-ideal system to study excitonic effects and light matter coupling in solid state systems featuring very distinct, well separated excitonic and trionic resonances even at room temperature \cite{Cao2015a,MakEtAl2016}. This is a consequence of the reduced dielectric screening in atomic monolayers along with heavy exciton masses \cite{Chernikov2014}. Furthermore, they are flexible yet robust, easy to fabricate, and they can be integrated in dielectric, metallic, semiconducting and even monolayer-based heterostructures. The coupled spin and valley physics lead to effects such as the valley Hall effect and the valley coupling to the optical helicity\cite{Xiao2012,Xu2014}. While devices such as light emitting diodes \cite{Ross2014,Baugher2014}, solar cells \cite{Pospischil2014}, ultra-fast photodetectors\cite{Lopez-Sanchez2013} and single-photon emitters\cite{He2015,Srivastava2015,Chakraborty2015,Koperski2015,Tonndorf2015} have been demonstrated already, the potential of TMDCs for fundamental studies of exciton -light coupling in the presence of quantum confinement is yet to be exploited. In this field, the behaviour of Coulomb-correlated carrier complexes in the presence of a carrier gas is of particular interest potentially featuring interesting effects such as the polarons and polariton formation at Fermi-edge singularities\cite{Sidler2016}, and even exciton-mediated superconductivity\cite{Cotlet2015,Laussy2010,Cherotchenko2016}. In order to investigate these effects, a profound understanding of the interaction between excitons, trions and free carriers is absolutely essential. This may include subtle energy shifts and renormalization of excitonic and trionic binding energies due to the presence of free carriers which has been observed in other materials systems \cite{Astakhov1999,Astakhov2002b,Astakhov2002}.
%
%
%
%
%

Experimentally, the influence of a free carrier gas on the excitonic response in monolayers of MoS$_2$ and WS$_2$ \cite{Chernikov2014,Mak2013} has been studied in absorption measurements close to the Mott transition in gated devices. Similar results have recently been confirmed in photoluminescence measurements on MoSe$_2$ using photodoping to create free carriers \cite{CadizEtAl2016}. Results on MoSe$_2$ show only slightest changes in trion dissociation energy on the sub-meV scale. Nevertheless, MoSe$_2$ monolayers are particularly promising for studies of such effects, since their emission spectra features a strong, spectrally well separated trion signal. On most substrates no defect related, potentially disturbing features are observed as compared to WSe$_2$ monolayers for example. This clear spectrum makes MoSe$_2$ monolayers well suitable to study the effect of excess carriers on these Coulomb-correlated complexes. 

Although, a change in trion dissociation energy has been observed in a subtle sub-meV range on MoSe$_2$, no quantitive theory has been provided for this behaviour so far. Here, we enhance the amount of photodoping on MoSe$_2$ monolayers at cryogenic temperatures using higher excitation powers and longer treatment times in a first step. Consequently, we can observe changes in the trion dissociation energy on the order of several meV. We find, that both the exciton and the trion binding energy in MoSe$_2$ sensibly depends on the excess carrier density in the monolayer. While previous investigations were rather focused on the origin of photo-doping process\cite{CadizEtAl2016}, we provide a newly developed model for the interaction between free carriers, excitons and trions. This model is based on the original variational approach that takes into account both screening and phase space filling effects. This model is in good quantitative agreement with our experimental findings.

\section*{Methods}

Monolayers of MoSe$_2$ were deposited onto 285 nm thermal oxide on Si wafers via conventional exfoliation from bulk crystals. The SiO$_2$ thickness was chosen to be 285 nm to improve the monolayer contrast in our optical microscope. First, the exfoliated monolayers were characterized using Raman and photoluminescence spectroscopy to check their optical quality. 
Photoluminescence spectra are recorded at 10 K, via exciting the monolayer non-resonantly with a frequency doubled Nd:YAG laser at 532 nm. The excitation laser is utilized to create excitons and trions in our monolayer as well as to activate additional carriers in the heterostructure. 

\section*{Experiment}

Fig. 1a) depicts a series of photoluminescence spectra of such a MoSe$_2$ monolayer under non-resonant photoexcitation at 532 nm. In this experiment, the intensity of the non-resonant pump laser is successively increased from 500 $\mu$W to 10 mW to study the evolution of the emission features which were previously identified as stemming from the excitonic (X) and the trionic carrier complex (X-) \cite{Mak2013}. Along with the increase of the peak intensity, the emission features are subject to a broadening of their emission linewidth, as depicted in fig 1b). The monotonous increase of the emission linewidth of both, the exciton as well as the trion signal can be related to a combination of effects. One effect could be an increased dephasing of the carrier complexes interacting with a successively increasing carrier gas in the background. In addition, the pump laser induces heat in our system, which results in an increasing excitation of phonons, which couple to our excitons and cause a broadening of the emission linewidth \cite{MoodyEtAl2015}. 

We will now comparatively monitor the power-dependent emission features of the X and the X- complex in our sample. Thus, we re-plot the series of emission spectra shown in fig. 1a) by normalizing each spectrum to the excitonic signal, as shown in fig 2a). This procedure reveals the increasing dominance of the trionic feature towards higher excitation powers. This already indicates that the non-resonant pump laser activates additional electrons in our sample, which contribute to the formation of trionic complexes. Most likely, this activation stems from optical ionization of trap centers in the substrate or in the substrate-monolayer interface, which is followed by a transfer of the free charge carrier into the monolayer\cite{CadizEtAl2016}.

A more controlled method to tune the electron density without modifying the excitonic density in our monolayer is shown in fig2 b): Conveniently, the photoexcitation process of additional free carriers takes place on a very slow timescale (seconds to minutes). Thus, the fraction of excitons versus trions can be tuned simply by adjusting the illumination time of the monolayer. Fig. 2b) depicts a series of spectra where the exication laser 5 mW is kept constant, and each spectrum is recorded after 1 minute.
More quantitatively, we plot the ratio of the integrated intensities of the trion and exciton emission peaks as functions of the exposure time in fig. 2c), for various excitation powers. Starting from a low excitation power, we are in a regime where the exciton signal is still dominating the spectrum. However, even with constant power, the X-/X fraction is already monotonously increasing with time, exhibiting a saturation behaviour. This indicates a self-liminting activation process, typical for the succesivel activation of finit number of defect states. An increase in the pump power then leads to a further increase of this ratio, which serves as a proportional measure for the number of electrons per exciton in our system\cite{CadizEtAl2016}. Within the pump power range of 0.5 mW to 10 mW (power measured in front of our microscope objective) and an exposure time up to 60 minutes, we manage to tune the X-/X fraction between a number as low as 0.3 up to 2.2. This gives us a convenient basis to study electron-hole correlation energies in the presence of a free electron gas. 
We thus investigate the emission energies of both optical resonances as a function of both the pump power and the exposure time, which are directly correclated to the excess electron density. 

We extract the peak energies of the X and the X- resonance by fitting the spectra to a double Gaussian envelope. Due to the almost background-free spectra, the energies of both resonances can be extracted with a great accuracy (100$\mu$eV). The evolution of the energies of both the exciton and the trion as functions of the exposure time are subject to a step-wise increase of the pump power shown in fig. 3a). As the excitation power is increased, both the X and the X- signal experience a similar red-shift, resulting from the power-induced renormalization of the bandgap of the monolayer. The contribution of the microscopic processes of this renormalization are not fully determined (dielectric screening vs. sample heating), however are of a minor importance in our study, since the energy scale is approximately two orders of magnitude smaller than the correlation energies of our excitonic complexes. A second effect, which can be directly observed in fig. 3a), is the shift of both the X and X- signal as the pump laser intensity is kept at the same level. Here, the behaviour of both resonances is fundamentally different: as the pump intensity is kept constant and charges successively accumulate in the monolayer, the X feature is subject to a continuous blueshift in energy, whereas the X- feature is redshifted by a similar magnitude (hundreds of $\mu$eV up to meV). The blueshift of the exciton in the presence of an accumulating electron gas is a strong signature of a renormalization of the exciton binding energy. In the presence of additional screening, Coulomb coupling between electrons and holes is reduced \cite{Edelstein1989,Kleinmann,Bigenwald2001}, and the excitonic emission energy eventually approaches the free carrier bandgap. Due to the modest pump powers, the shift of the exciton in our experiment is significantly smaller than the exciton binding energy. We thus conclude that both the exciton density and the free carrier density are significantly lower than the Mott density (ranging around $10^{
14}$ carriers /cm$^2$) in the  MoSe$_2$ monolayer. 

The redshift of the energy of the trion, on the other hand, is a strong indication that the trion binding energy is enhanced by the electron reservoir, which over-compensates the screening induced blueshift of the exciton. In order to directly correlate the trion binding energy with the approximate number of excess carriers per exciton, we plot the energy ($E_{X}-E_{X-}$) as a function of the X-/X ratio in fig. 3b). As we have discussed above, the fact that we are significantly below the Mott transition in our material justifies this representation which disregards the absolute number of excitons and trions in our sample, since we can exclude exciton-exciton scattering processes to play a dominant role in this pumping regime. 

We find that the trion dissociation energy in our system is a linear function of the intensity ratio between X- and X, and thus of the number of electrons in our monolayer. The fact that this linear behavior is furthermore present throughout the stepwise increases in the pump power additionally justifies our interpretation, and evidences that the interaction of excitons and trions with free carriers is the main reason of these energy shifts.

\section*{Theory}

In order to explain the monotonous linear increase of trion binding energy with the increase of free carrier concentration we use the variational approach. 
The Schroedinger equation for the wave-function of the electron-hole 
relative motion in the plane of the layer $\Psi (\rho)$ in cylindrical coordinates reads:
\begin{equation}\label{1}
\left[\frac{-\hbar^2}{2\mu}\frac{1}{\rho}\frac{\partial}{\partial\rho}\left(\rho\frac{\partial}{\partial\rho}\right) - \int_0^\infty{J_0(k\rho)V_Ckdk-E_{ex}}\right]\Psi(\rho)=0 , 
\end{equation}
where $\mu=\frac{m_{e}m_{h}}{m_{e}+m_{h}}$ is the reduced mass of electron-hole relative motion, $J_{0}(k\rho)$ is the zeroth order Bessel function, $V_C$ is the screened Coulomb potential. 

Here we describe the free carriers as a degenerate Fermi gas at zero temperature. We account for the exclusion effect associated with the phase space filling. All electronic states below the Fermi level are assumed to be occupied, which is why they cannot contribute to the exciton state. Taking into account this exclusion effect the trial function of the exciton can be written in the form \cite{Pikus1992}: 
\begin{align}
\Psi(\rho)=&\int_0^{\infty}{J_{0}(k\rho)f(k)kdk},	\\
f(k)=&\frac{B}{(A+k^2)^{3/2}}\Theta(k-k_F) \nonumber
\end{align}
where $\Theta(k-k_F)$ is the Heaviside function,$k_F=\sqrt{2\pi n_{2D}}$, $n_{2D}$ is the density of free carriers, $A$ is a variational parameter, inversely proportional to squared exciton Bohr radius. $B$ can be found from the normalization condition:
\begin{equation}\label{3}
\int_0^{\infty}\Psi(\rho)^22\pi\rho d\rho=1 
\end{equation}
In the limiting case of $k_F$=0 the introduced wave function reduces to the hydrogen-like wave-function: 
\begin{equation}
\Psi(\rho)=\sqrt{\frac{2}{\pi}}\frac{1}{a}exp(-\frac{\rho}{a}),\nonumber
\end{equation}
where $a$ is the exciton Bohr radius.

Varying the parameter $A$ the exciton binding energy can be found as the minimum value of $E_{ex}$, where 
\begin{equation}\label{4}
E_{ex}=\int_0^{\infty}\left[\frac{-\hbar^2}{2\mu}\frac{1}{\rho}\frac{\partial}{\partial\rho}\left(\rho\frac{\partial}{\partial\rho}\right) - \int_0^\infty{J_0(k\rho)V_Ckdk}\right]\Psi(\rho)^{2}2\pi\rho d\rho
\end{equation}

The screened Coulomb potential in eq.\eqref{1} reads \cite{Klochikhin2011,Klochikhin2014}:
\begin{equation}\label{5}
V_C=\frac{e^2}{\epsilon k(1+(\kappa/k)(1-\Theta(k-2k_F)\sqrt{1-(2k_F/k)^2}))}
\end{equation}
Here $\epsilon$ is the mean dielectric constant that takes into account the substrate, $\kappa=\frac{2g_{\nu}m_{e}e^2}{\epsilon\hbar^2}$ is the screening constant, $g_{\nu}$ is the valley degeneracy factor. The electron-density dependened potential
(5) is suitable for the description of the dependence of the exciton binding energy on the free
career concentrations. Its advantages over the frequently used Yukawa potential are addressed in 
\cite{Klochikhin2011,Klochikhin2014}

For the trion case we simplify the problem and assume that the two electrons (we consider the X$^-$ case without loss of generality) are in the singlet state so that they are characterized by orthogonal spin functions and identical spatial wave functions. The trion binding
energy may be found as the solution of the Coulomb problem with a hole of charge
$+e$ and mass $m_h$ and an electron pair of charge  $-2e$ and mass $2m_e$.
In this case the trion wave function can be expressed as a sum of two parts, corresponding to the electrons composed by the states that lie below and above the Fermi level respectively:
\begin{align}
\Psi(\rho)&=\int_0^{\infty}{J_{0}(k\rho)f_{tr}(k)kdk},	\\
f_{tr}(k)&=\frac{B}{(A+k^2)^{3/2}}\Theta(k-k_F)+\frac{C}{(D+k^2)^{3/2}}\Theta(k_F-k), \nonumber
\end{align}
where $A$ and $D$ are variational parameters and $B$, $C$ can be found from normalization conditions. Here we account for the exclusion principle for both electrons: the photoexcited electron can only be formed by
free states, while the resident electron can only be taken from the states below the
Fermi level. The good estimate for the trion binding energy is found in a similar manner as in \eqref{4} by minimization over both variational parameters $A$ and $D$. 

The described method was used to estimate exciton and trion binding energies in a MoSe$_2$ monolayer. We have used the following parameters: $m_{ex}$=0.62$m_0$, $\epsilon$=11 that averages monolayer dielectric constant \cite{Ouyang2016} and substrate dielectric constant ($\epsilon$=1.5). In the absence of free carriers, this model yields the exciton binding energy of about $E$=300 meV.
The results for exciton and trion binding energies at different electron densities calculated with the described model are shown in fig. 3(c). The binding energies with the presence of free carriers calculated with this approach are somewhat lower than the values for MoSe$_2$. This might be improved by taking the screening more accurately e.g. using Keldysh model\cite{Keldysh}. The approach developed here has an advantage of the utmost simplicity that allows tracing
the main dependencies quasi-analitically. It correctly describes the experimental data in what concerns
the dependence of the energy splitting between exciton and trion resonances on the 
free carrier density. It can be seen that at low densities (3 orders of magnitude below the Mott density) the energy difference increases linearly with electron density.

\section*{Conclusions}

We have investigated the Coulomb correlation energies of photo-excited excitons and trions in monolayers of MoSe$_2$ at cryogenic temperatures. By taking advantage of an inherent photoexcitation of additional charge carriers in our structure, we were able to tune the ratio between exciton and trion peak intensities by a factor of 7. We observe a continuous screening of the exciton correlation on the meV scale, and more importantly, a simultaneous increase of the trion binding energy up to several meV as the electron density is increased. These experimental findings are reproduced with the variational calculation that accounts for the screening and phase space filling effects for excitons and trions.
We believe that these findings are of fundamental interest in the emerging research field devoted to Fermi-Bose mixtures, in particular targeting to explore interactions of bosonic excitations in the presence of a Fermi-sea.  Understanding subtle power-depending signatures of the primary resonances in monolayers of TMDCs will help harnessing electron-exciton coupling to induce superconductivity mediated by excitons. 

\section*{Acknowledgement}

We acknowledge S. Tongay for his assistance in sample preparation.


%
\section{Figures}

\begin{figure}[ht]
\centering
\includegraphics[width=\linewidth]{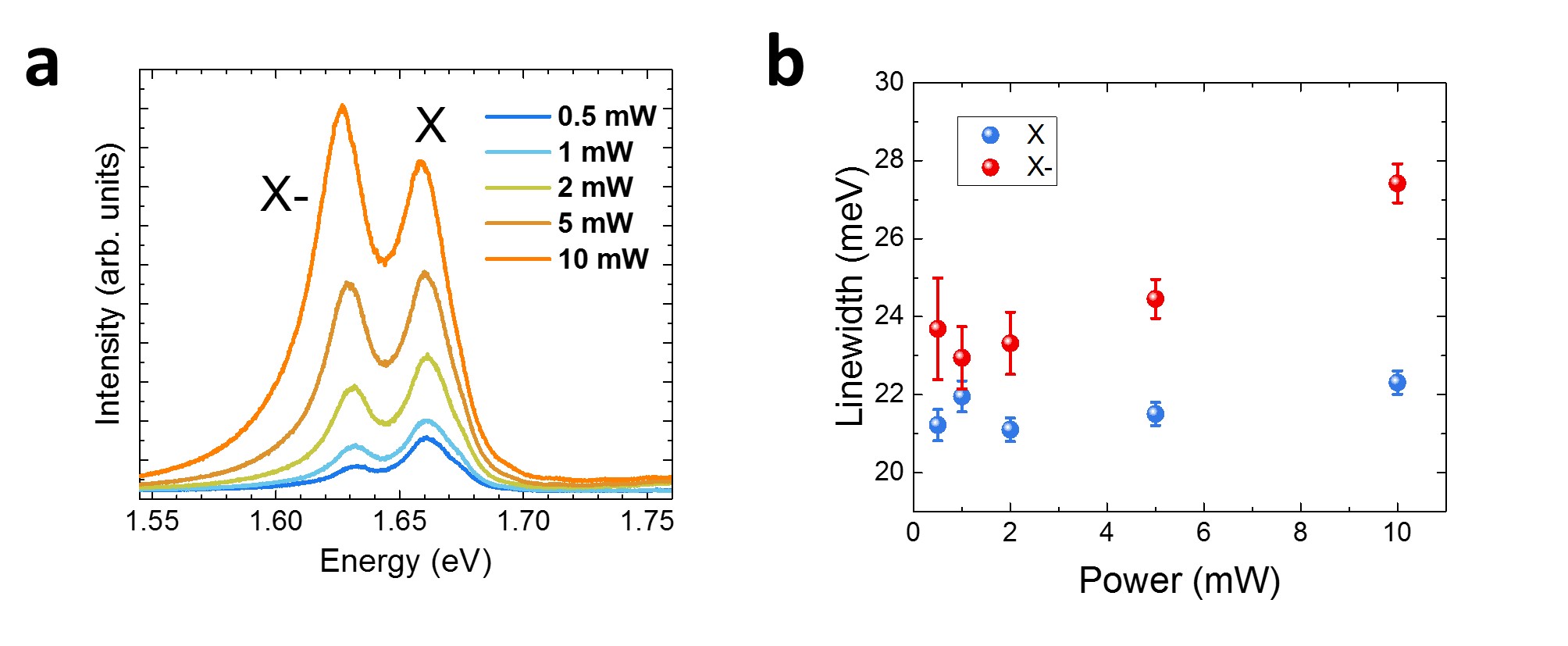}
\caption{a) Power series of MoSe$_2$ monolayer at 5K under 532 nm excitation b) Corresponding linewidths of excitonic and trionic resonances}
\end{figure}
\begin{figure}[ht]
\centering
\includegraphics[width=\linewidth]{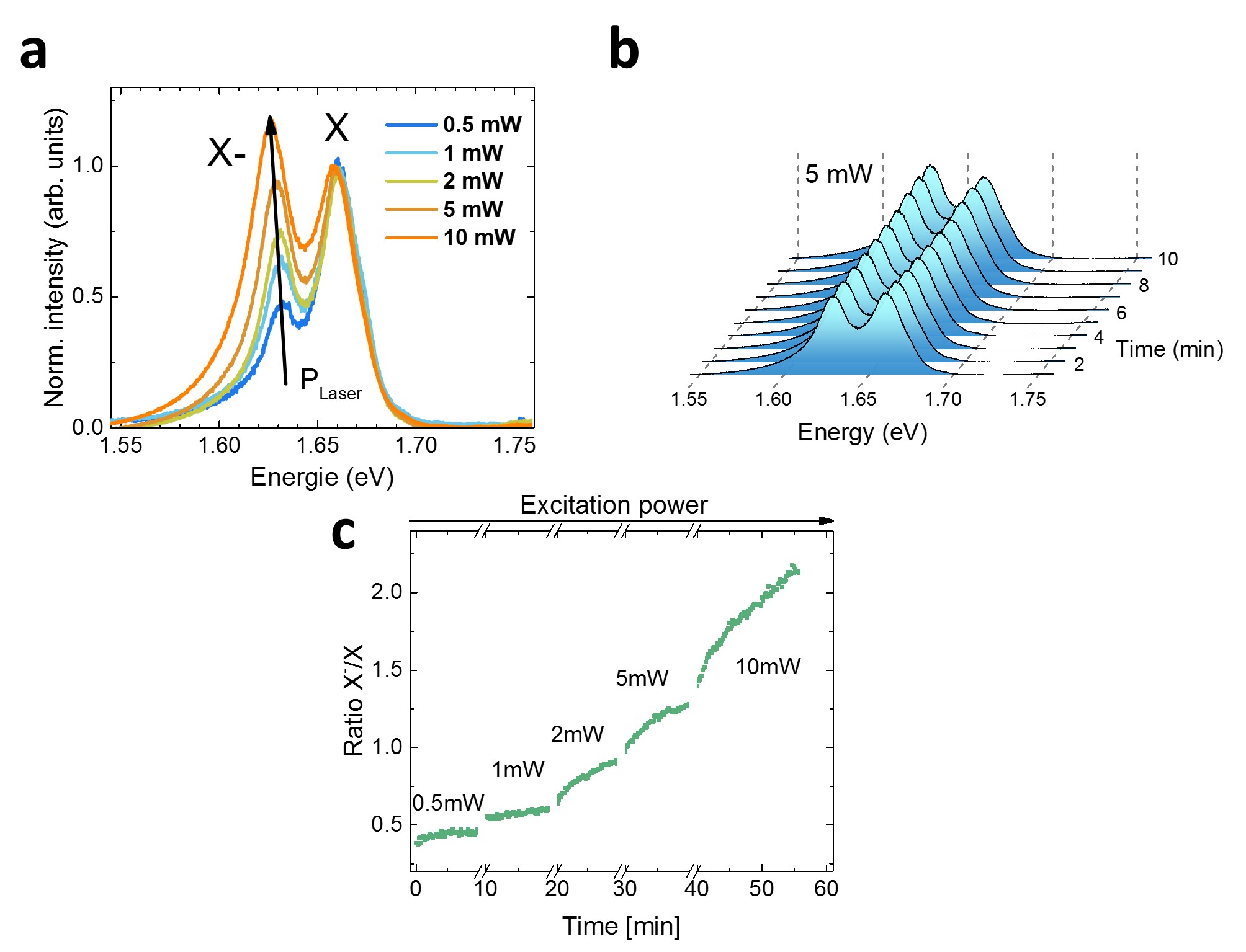}
\caption{a) Power series of hte PL spectra normalized to the excitonic resonance b) Time evolution of PL spectra under constant exciation power of 5 mW  c) Ratio of integrated peak intensities as a function of time and excitation power}
\end{figure}
\begin{figure}[ht]
\centering
\includegraphics[width=\linewidth]{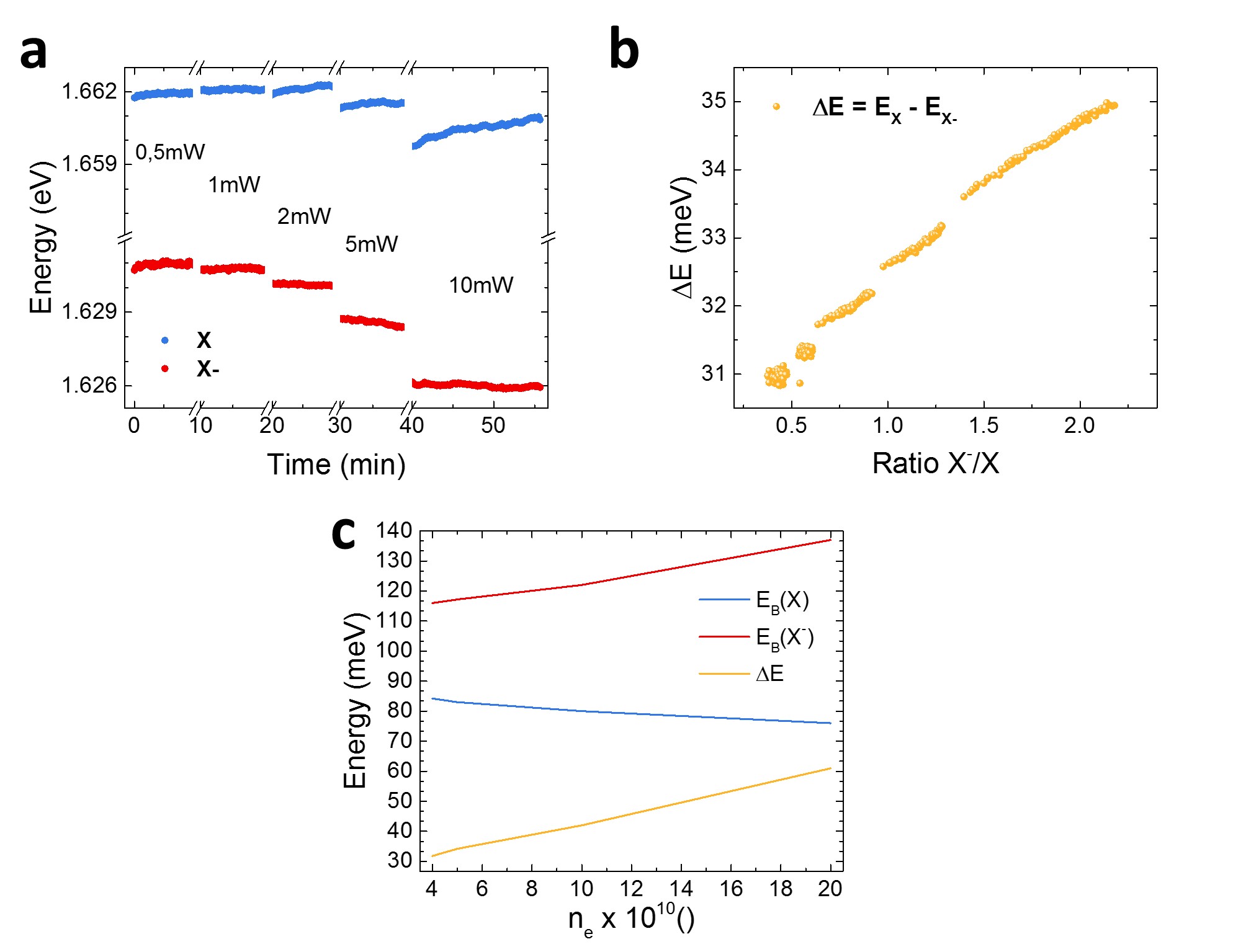}
\caption{a) Temporal evolution of excition and trion energy b)Trion dissociation energy as a function of X-/X ratio c) Calculated binding energies modelled with the variational approach and the corresponding difference}
\end{figure}


\end{document}